\documentclass{article}
\usepackage[pdftitle={Multiscale neuronal modeling},
  pdfauthor={Bauer, Engblom, Mikulovic, Senek},
  pdffitwindow=true,
  breaklinks=true,
  colorlinks=true,
  urlcolor=blue,
  linkcolor=red,
  bookmarks=false,
  citecolor=red,
  anchorcolor=red]{hyperref}
\usepackage[utf8]{inputenc}
\usepackage[english]{babel}
\usepackage{graphicx,color}
\usepackage{amsmath}
\usepackage{amssymb}
\usepackage{mathtools}
\usepackage{bbm}
\usepackage{marvosym}
\usepackage{epstopdf}
\usepackage{algorithm}
\usepackage{algorithmic}
\usepackage{verbatim}
\usepackage{paralist}
\usepackage{chemfig}
\usepackage[numbers,sort]{natbib}

\definecolor{darkgreen}{rgb}{0,0.5,0}

\newcommand{\X}{\mathbb{X}}
\newcommand{\Stoich}{\mathbb{S}}
\newcommand{\fatmu}{\boldsymbol{\mu}}

\newcommand{\Expect}{\mathbb{E}}

\newcommand{\Mstates}{M_{\mbox{\scriptsize{states}}}}
\newcommand{\Ntransitions}{N_{\mbox{\scriptsize{transitions}}}}
\newcommand{\Mcompartments}{M_{\mbox{\scriptsize{compartments}}}}

\newcommand{\fatJ}{\mathbf{J}}
\newcommand{\fatE}{\mathbf{E}}

\newcommand{\fatD}{\mathbf{D}}

\newcommand{\review}[1]{#1}

\numberwithin{equation}{section}
\numberwithin{table}{section}
\numberwithin{figure}{section}

\begin{document}

\title{Multiscale modeling via split-step methods in neural firing}

\author{Pavol Bauer \thanks{Pavol Bauer, Stefan Engblom, and
    Aleksandar Senek are with the Division of Scientific Computing,
    Department of Information Technology, Uppsala University, Box 337,
    SE-751 05 Uppsala, Sweden, e-mail:
    \href{mailto:pavol.bauer@it.uu.se}{pavol.bauer@it.uu.se},
    \href{mailto:stefane@it.uu.se}{stefane@it.uu.se},
    \href{mailto:Aleksandar.Senek.1902@student.uu.se}{Aleksandar.Senek.1902@student.uu.se}.}
  \and Stefan Engblom \footnotemark[1] \thanks{Corresponding author,
    address as above. Office phone +46 18 471 27 54, fax +46 18 51 19
    25.} \and Sanja Mikulovic \thanks{Sanja Mikulovic, is with the
    Department of Neuroscience, Biomedical Centrum, Uppsala
    University, Box 593, SE-751 24 Uppsala, Sweden, e-mail:
    \href{mailto:sanja.mikulovic@neuro.uu.se}{sanja.mikulovic@neuro.uu.se}.}
  \and Aleksandar Senek \footnotemark[1]}

\maketitle

\begin{abstract}
  Neuronal models based on the Hodgkin-Huxley equation form a
  fundamental framework in the field of computational
  neuroscience. While the neuronal state is often modeled
  deterministically, experimental recordings show stochastic
  fluctuations, presumably driven by molecular noise from the
  underlying microphysical conditions. In turn, the firing of
  individual neurons gives rise to an electric field in extracellular
  space, also thought to affect the firing pattern of nearby neurons.

  We develop a multiscale model which combines a stochastic ion
  channel gating process taking place on the neuronal membrane,
  together with the propagation of an action potential along the
  neuronal structure. We also devise a numerical method relying on a
  split-step strategy which effectively couples these two processes
  and we experimentally test the feasibility of this approach. We
  finally also explain how the approach can be extended with Maxwell's
  equations to allow the potential to be propagated in extracellular
  space.

\medskip


\medskip



\end{abstract}


\section{Introduction}
\label{sec:intro}

Neurons are responsible for encoding information in the central
nervous system. Lower level functions are many times gathered at
specific parts of the brain, processing information received from
neurons throughout the body, often by means of signaling nerves, which
are bundles of axons that reach out from the neurons. Higher level
functions depend on remarkably larger and complex networks of neurons with
various types of feedback loops.

The chemical connections between neurons are handled via synapses,
where neurotransmitters are extruded into the extracellular space by
the presynaptic neuron. The neurotransmitters form a part of a
chemical process which may initiate a potential wave into the
postsynaptic neuron --- this wave is known as the \emph{action
  potential}. During the action potential, the membrane potential
quickly rises and falls, and the resulting signal propagates along the
cell \cite{purves_neuroscience_2012}. The process that underlies this
propagation is the regulation of ion concentrations in both the
intracellular cytoplasm and the extracellular space caused by integral
membrane proteins called \emph{ion channels}.

There are many variants of ion channel proteins, whose functions are
only recently better understood through studies using experimental
techniques such as X-ray crystallography
\cite{rees_crystallographic_2000}. To our current understanding, ion
channels reside at one of many conformal states, which are either
\emph{closed} (non-conducting) or \emph{open} (conducting). If the
conformal state is open, ions are allowed to pass through pathways
called pores from the extra- to the intracellular space, or in the
opposite direction. If the conformal state is closed, ions are blocked
from entering the channel \cite{hille}.  Ion channels become activated
either in response to a chemical ligand binding, or in response to
voltage changes on the membrane \cite{hille,purves_neuroscience_2012},
so-called voltage-gated ion channels.  In this work, we focus on
voltage-gated ion channels, which are important for the initiation and
the propagation of action potentials along the neuronal fiber.

Mathematical models of neurons were initially formed by measuring
electrically induced responses of a neuron using techniques such as
the voltage- or current clamp. From those experiments, transition
properties and parameters of single channel gating could be identified
\cite{hille}.

Historically, models of firing neurons were formulated as ordinary
differential equations (ODEs). \review{However, in some specific cases
  experimental as well as theoretical findings suggest that the gating
  of ion channels is more accurately described by a stochastic
  process.} The variance of the gating process, also known as the
\emph{channel noise}, is thought to be important for information
processing in the dendrites and explain different phenomena regarding
action potential initiation and propagation \cite{schneidman_ion_1998,
  white_channel_2000,faisal_noise_2008, cannon_stochastic_2010}.
\review{For example, it has been shown that intrinsic noise is
  essential for the existence of subthreshold oscillations in stellate
  cells \cite{dorval_channel_2005}, that it contributes to irregular
  firing of cortical interneurons \cite{stiefel_origin_2013}, and that
  it can explain firing correlation in auditory nerves
  \cite{moezzi_ion_2016}.}

Another important aspect of neuronal modeling is the investigation of
action potentials propagating outward the neuron into the
extracellular space.  Simulations of such extracellular fields is one
of the important methods used in computational neuroscience
\cite{einevoll_modelling_2013}. Common usage includes the validation
of experimental methods such as EEG and extracellular spike
recordings, or in the modeling of physiological phenomena which can
not be easily investigated empirically \cite{einevoll_modeling_2010}.

In this paper we present a novel three-stage multiscale
model\review{ing framework} consisting of the following components:
\begin{inparaenum}[(a)]
\item on the microscale, the gating process of the ion channels is
  governed by a continuous-time discrete-state Markov chain,
\item on the intermediate scale, the current-balance and the cable
  equation which are responsible for the action potential initiation
  and propagation is integrated in time as an ODE,
\item on the macroscale, the propagation of the trans-membrane
  current into an electrical field in extracellular space is achieved
  using partial differential equations (PDEs).
\end{inparaenum}

In \S\ref{sec:scales} the three modeling layers are explained in some
detail. The numerical method via split-step modeling is summarized in
\S\ref{sec:num}, where we also explain how the spatial coupling is
achieved efficiently. We illustrate our method by some relevant
examples in \S\ref{sec:experiments} and offer a concluding discussion
in \S\ref{sec:conclusions}.


\section{Neuronal modeling at different scales}
\label{sec:scales}

\begin{figure}[H]
  \centering
  \includegraphics[width=1\textwidth]{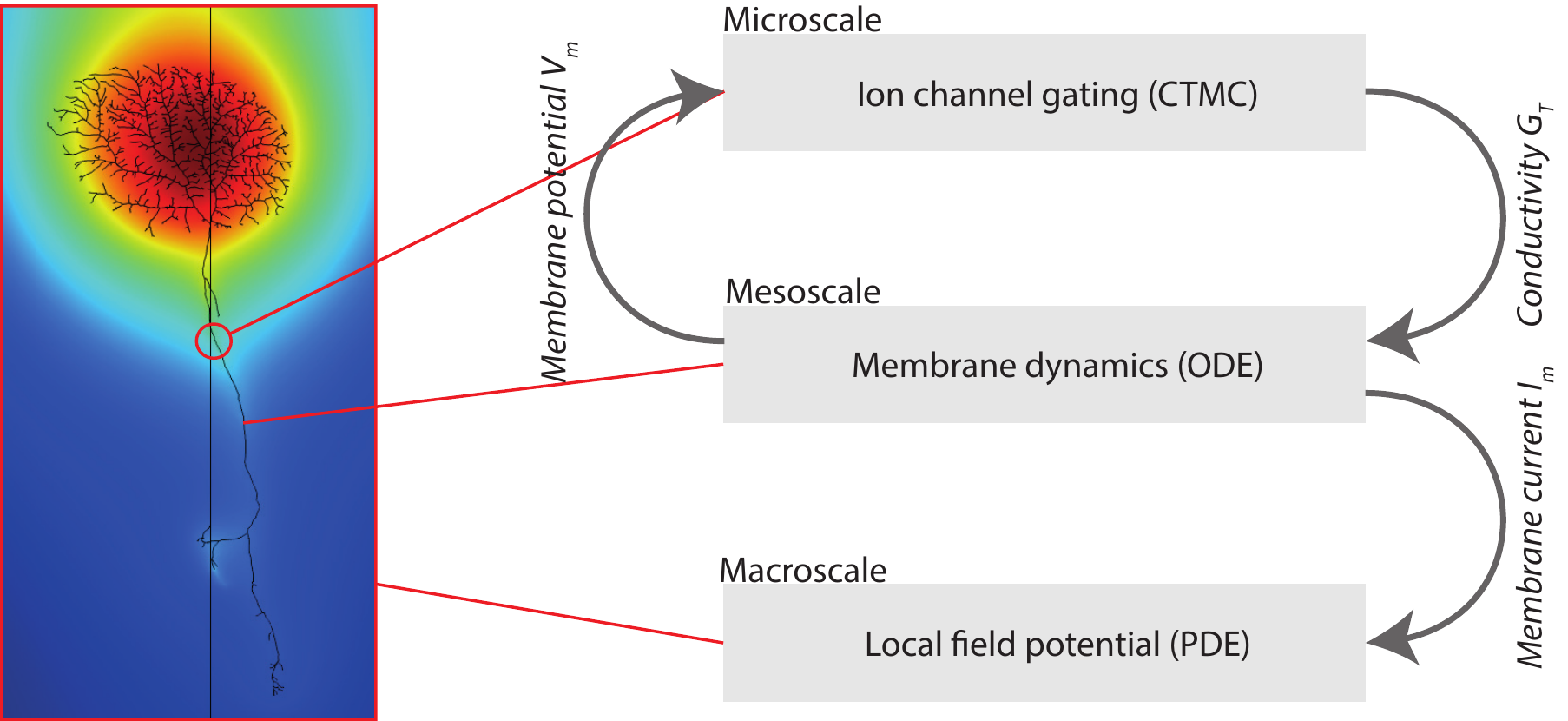}
  \caption{\review{Schematic overview of the proposed multiscale
      modeling framework. The CTMC solution on the microscale depends
      on the membrane voltage that is computed on the mesoscale. This
      coupling is bidirectional; the mesoscale solution depends on the
      microscopic stochastic ion current. The macroscopic solution
      considered here depends solely on the mesoscopic membrane
      current.}}
  \label{fig:scheme_overview}
\end{figure}

We now describe the modeling \review{framework} at the individual
scales, including the associated modeling assumptions. The microscale
physics in the form of continuous-time Markov chains (CTMC)
\review{for the ion channels and the associated ion currents} are
discussed in \S\ref{sec:microscale}, the ODE-model for the action
potential \review{along the neuronal geometry} in
\S\ref{sec:mesoscale}, and the PDE-model of the extracellular electric
field via Maxwell's equations in \S\ref{sec:macroscale}. For
convenience, a schematic explanation of the modeling framework is
found in Figure~\ref{fig:scheme_overview}.

\subsection{Microscale: ion channel gating}
\label{sec:microscale}

The gating of voltage-dependent ion-channels is modeled by a Markov
process. Hodgkin and Huxley \cite{hodgkin_quantitative_1952} first
proposed that the gating is governed by a set of gating variables.
We assume that each gating variable takes values in a
discrete state space and that a single combination of gating variables
corresponds to an open and conducting ion channel \cite{hille}.

\begin{figure}
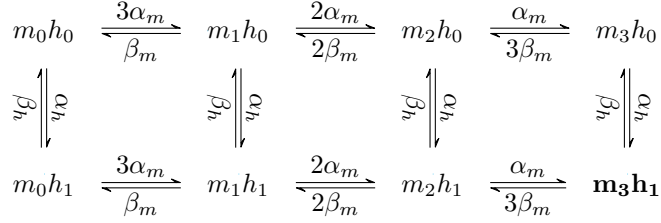

  \centering
  \schemestart 
  $m_0h_0$ \arrow(A--B){<=>[$3 \alpha_m$][$\beta_m$]} 
  $m_1h_0$ \arrow(B--C){<=>[$2 \alpha_m$][$2\beta_m$]} 
  $m_2h_0$ \arrow(C--D){<=>[$\alpha_m$][$3\beta_m$]} $m_3h_0$
  \arrow(@A--E){<=>[$\alpha_h$][$\beta_h$]}[-90] $m_0h_1$
  \arrow(@B--F){<=>[$\alpha_h$][$\beta_h$]}[-90] $m_1h_1$
  \arrow(@C--G){<=>[$\alpha_h$][$\beta_h$]}[-90] $m_2h_1$
  \arrow(@D--H){<=>[$\alpha_h$][$\beta_h$]}[-90] $\mathbf{m_3h_1}$
  \arrow(@E--@F){<=>[$3 \alpha_m$][$\beta_m$]}
  \arrow(@F--@G){<=>[$2 \alpha_m$][$2 \beta_m$]}
  \arrow(@G--@H){<=>[$\alpha_m$][$3\beta_m$]}
  \schemestop
  \caption{Kinetic scheme for the $m_3h_1$ sodium channel gating
    proposed by Hodgkin and Huxley
    \cite{hodgkin_quantitative_1952}. Only the gating state $m_3h_1$
    represents an open ion channel, for all other states the ion
    channel is closed.}
  \label{fig:scheme}
\end{figure}

The gating states and the transitions between them can be written in
the form of a kinetic scheme. An example for the voltage-gated
$m_3h_1$ sodium channel is depicted in Figure~\ref{fig:scheme}.
\review{The notation for the channel states indicates that there are
  two involved gating variables $m$ and $h$, where variable $m$ takes
  one of four states and, independently, variable $h$ takes one of two
  states. In total we thus arrive at 8 states with a total of 10
  reversible transitions, see Figure~\ref{fig:scheme}. The total
  number of channels in the open state $m_3h_1$ implies a certain
  conductivity as will be detailed below.} The transition rates
between the states depend on the membrane voltage and on the state
itself. When these transitions take place in a microscopic environment
where molecular noise is present, a continuous-time Markov chain
(CTMC) is the most suitable model.

\begin{figure}[htb!]
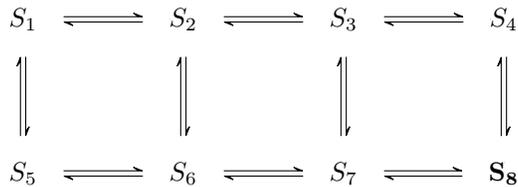

  \centering
  \schemestart 
  $S_1$ \arrow(A--B){<=>} 
  $S_2$ \arrow(B--C){<=>} 
  $S_3$ \arrow(C--D){<=>} $S_4$
  \arrow(@A--E){<=>}[-90] $S_5$
  \arrow(@B--F){<=>}[-90] $S_6$
  \arrow(@C--G){<=>}[-90] $S_7$
  \arrow(@D--H){<=>}[-90] $\mathbf{S_8}$
  \arrow(@E--@F){<=>}
  \arrow(@F--@G){<=>}
  \arrow(@G--@H){<=>}
  \schemestop
  \caption{An equivalent scheme to Figure~\ref{fig:scheme}, where each
    combination of gating variables corresponds to a discrete state of
    a Markov chain.}
  \label{fig:simple_scheme}
\end{figure}

We rewrite such schemes in a general, yet more manageable notation as
follows. Assign the different combinations of gating variables an
individual state $S_i$, $i = 1,2,\ldots,\Mstates$. For the sodium
scheme in Figure~\ref{fig:scheme} there are 8 states and only the
state $S_8$ corresponds to an open and conducting ion channel. The
resulting equivalent scheme is shown in Figure~\ref{fig:simple_scheme}
and an exemplary set of transitions and rates is
\begin{align}
  \label{eq:S3toS4}
  S_3 &\longrightarrow S_4 \text{, rate: } \alpha_m(V_m), \\
  S_3 &\longrightarrow S_2 \text{, rate: } 2\beta_m(V_m), \\
  S_3 &\longrightarrow S_7 \text{, rate: } \alpha_h(V_m),
\end{align}
where $V_m$ is the membrane potential at the location of the ion
channel. For this particular example, \review{the transition rates
  used by Hodgkin and Huxley are \cite{hodgkin_quantitative_1952}}
\begin{align}
\alpha_m(V_m) &= \frac{0.1(V_m+40)}{1-e^{-(V_m+40)/10}}, \label{eq:rate1} \\
\beta_m(V_m) &= 4e^{-(V_m+65)/18}, \label{eq:rate3} \\
\alpha_h(V_m) &= 0.07e^{-(V_m+65)/20},  \\
\beta_h(V_m) &= \frac{1}{1+e^{-(V_m+35)/10}}.\label{eq:rate2}
\end{align}

\review{The rates \eqref{eq:rate1}--\eqref{eq:rate2} were originally
  obtained by empirical fitting to experimental data and are
  formulated relative to the resting potential of the particular
  neuron model. We consider a small neural compartment $a$, obtained
  from a discretization of the neuronal fiber into small enough
  segments such that the potential is approximately constant within
  each compartment. Let there be} a total of $N^a$ ion channels of the
considered type (e.g.~the $m_3h_1$ sodium channel) on the
compartmental surface. At any instant in time $t$, let $s_i$ be the
number of channels in state $S_i$, $i = 1,2,\ldots,\Mstates$. The
scheme in Figure~\ref{fig:simple_scheme} now directly translates into
Markovian transition rules for the states $s \in \mathbf{Z}_+^8$,
\review{i.e.~the ion-channel counts}, thus defining a CTMC $(t,s) \in
\mathbf{R}_+ \times \mathbf{Z}_+^8$.

To model the electrophysiological properties of the ion channel, we
define \review{the single open channel conductance $\gamma$}, as well
as the neuronal membrane area $A^a$ such that the area density of ion
channels is $\varrho^a = N^a/A^a$. If $O^a$ is the count of the open
ion channels, e.g., for the above example this is $O^a = S^a_8$,
\review{we arrive at the total conductance}
\begin{align}
  G^a = \frac{O^a}{N^a} \varrho^a A^a \gamma.
\label{eq:stochcurrent}
\end{align}
Next, the trans-membrane current produced by the ion channel is given
by Ohm's law,
\begin{align}
  I^a = G^a (V^a_m - E),
\label{eq:ioncurrent}
\end{align}
where $E$ is the reverse potential of the ion channel, that is, the
membrane potential under which the trans-membrane current $I^a$ is
zero, \review{and where $V^a_m$ is the membrane potential in
  compartment $a$.}

In the macroscopic setting proposed by Hodgkin and Huxley
\cite{hodgkin_quantitative_1952}, the transitions between the gating
states happened according to first order reaction kinetics. The
fraction of closed channels $(1-o^a)$ becomes open with rate $\alpha$
and the fraction of open channels $o^a$ becomes closed with rate
$\beta$. \review{Note that $\alpha$ and $\beta$ may be
  voltage-dependent for specific types of ion-channels, as in
  \eqref{eq:S3toS4}--\eqref{eq:rate2}.}  The resulting ODE for each
macroscopic gating variable $o^a$ can then be written as
\begin{align}
  \frac{do^a}{dt} = \alpha (1 - o^a) - \beta o^a.
\end{align}

Under the macroscopic formulation, the ratio of open channels is
approximated as
\begin{align}
  \label{eq:macroion}
  \frac{O^a}{N^a} \approx \prod_{i=1}^{M}\limits [o^a_i]^{p_i},
\end{align}
\review{where $M$ is the number of gating particles for the particular
  ion channel model, and $p_i$ is the exponent of the $i$th gating
  variable $o_i$.}

\review{As a concrete example, the sodium channel $m_3h_1$ depicted in
  Figure~\ref{fig:scheme} is gated by gating variables $m$ and $h$,
  which in the macroscopic formulation have an exponent of three and
  one, respectively.  In relation to \eqref{eq:macroion} this means
  that $o_1=m$ with $p_1=3$, and $o_2=h$ with $p_2=1$, where $m$ and
  $h$ are here just symbols denoting the concentration of the gating
  variables.} The ratio of open channels is then approximated as
\begin{align}
  \frac{O_{Na}}{N_{Na}} \approx m^3 h^1,
\label{eq:detratio}
\end{align}
and the deterministic gating function for each variable obeys
\begin{align}
\label{eq:detgating1}
  \frac{dm}{dt} &= \alpha_{m} (1 - m) - \beta_{m} m, \\
  \frac{dh}{dt} &= \alpha_{h} (1 - h) - \beta_{h} h,
\label{eq:detgating2}
\end{align}
where $\alpha$ and $\beta$ are the voltage dependent transition rates
defined in \eqref{eq:rate1}--\eqref{eq:rate2}. \review{Solving
  \eqref{eq:detgating1}--\eqref{eq:detgating2}} and using
\eqref{eq:ioncurrent} thus constitutes the classical Hodgkin-Huxley
ODE model for the ionic current. By rather evolving the Markov chain
defined implicitly in Figure~\ref{fig:simple_scheme} and using
\eqref{eq:stochcurrent}--\eqref{eq:ioncurrent}, a stochastic current
is defined. \review{This stochastic model is the result of the
  microphysical assumption of discrete ion states obeying Markovian
  (memoryless) transition rules. For large enough numbers of ion
  channels, a transition into the corresponding deterministic model
  can be expected under rather broad conditions
  \cite{Markovappr}. When a fine enough compartmentalization of a
  particular neuron is made, however, the stochastic model can be
  expected to be more realistic in the sense that the effects missing
  in a corresponding deterministic model can not be ignored.}

We now proceed to discuss the appropriate model for evolving the
resulting current, be it stochastic or deterministic, over the
neuronal morphology.

\subsection{Intermediate scale: current-balance and cable equation}
\label{sec:mesoscale}

On the intermediate scale, we solve for the current-balance equation
of the neuron, describing the relation between the membrane voltage
and the different current sources. We first take the ionic current
sources from the microscale into account. We next model the
capacitance of the membrane, a so-called passive neuronal property. As
the neuronal morphology is divided into several compartments we also
include a term for the propagation of voltage between the
compartments, resembling the cable equation.

\begin{figure}[htb!]
  \begin{center}
    \includegraphics[width=0.45\textwidth]{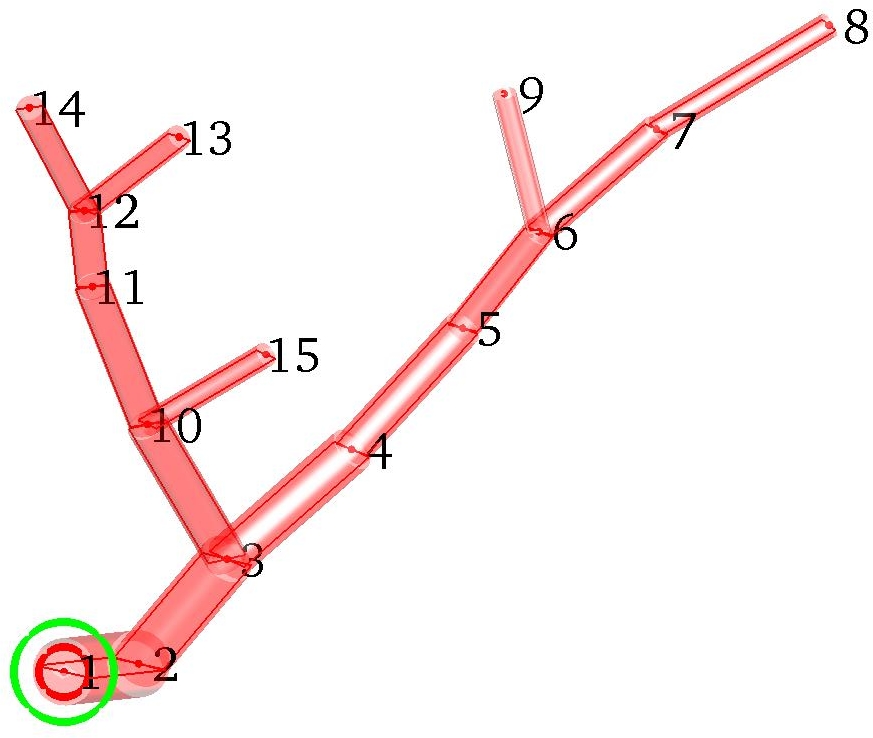}
    \includegraphics[width=0.45\textwidth]{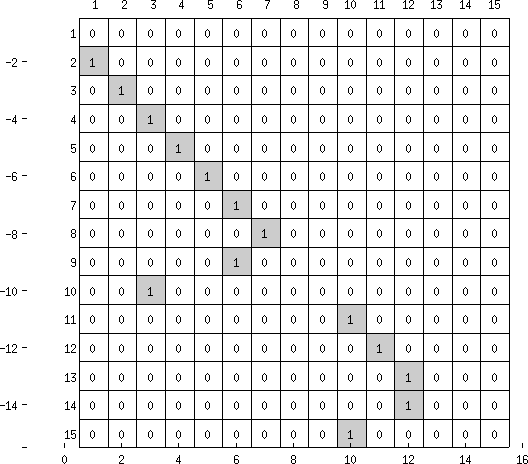}
  \end{center}
  \caption{Geometry of a neuron and the associated adjacency matrix
    $\mathcal{A}$ (figure adapted from~\cite{cuntz2010one}).
    \review{A nonzero entry in $\mathcal{A}$ implies a connection
      between the corresponding compartments.}}
  \label{fig:neuron_model}
\end{figure}

\review{As before,} denote the compartments by the index $a \in
\{1,\ldots,\Mcompartments\}$.  For each connection between
compartments, an entry is made in the adjacency matrix $\mathcal{A}$
\cite{cuntz2010one}. Typically, a compartment is connected to two or
three neighbors. If a compartment is connected to only a single
neighbor, it represents one of the end-points of the neuron.  An
example is found in Figure~\ref{fig:neuron_model}.

The current flowing in each compartment is composed from different
components.  The current source of compartment $a$ is the total axial
current with respect to its connected compartments
$b\in\mathcal{B}(a)$. \review{Note that as $\mathcal{A}$ represents a
  directed graph, the connected component $b\in \mathcal{B}(a)$
  represents in-neighbors that are connected via an edge in the
  direction to vertex $a$, as well as out-neighbors which are
  connected by an edge outwards of vertex $a$.}  According to Ohm's
law the current equals
\begin{align}
  \label{eq:conduct}
  I^a_{axial} = \sum_{b\in\mathcal{B}(a)} G^a_b(V^a_m - V^b_m) =
  \sum_{b\in\mathcal{B}(a)} G^a_b V^a_m - \sum_{b\in\mathcal{B}(a)} G^a_b V^b_m,
\end{align}
where $G^a_b$ is the conductance between compartments $a$ and $b$, and
where $V^a_m$ is the trans-membrane potential of compartment $a$.

We then add the ionic currents computed at the microscale.  For the
sake of generality we extend the previous notation of a single type of
ion channel into $c$ types, $\mathcal{T}=\{T^1,\ldots,T^c\}$, acting
simultaneously. Generalizing \eqref{eq:ioncurrent} we arrive at the
total ionic current
\begin{align}
  I^a_{ionic}=\sum_T G^a_T (V^a_m-E_T).
\label{eq:ion_middle}
\end{align}

\review{
We now deviate slightly from the discussion and first consider a
space-continuous solution obtained by solving for the propagation of
the potential on a line oriented along the $x$-axis. Adding the axial
current flux $I_{axial}$, the total trans-membrane current flux $I_m$,
and a possibly additional external current source $I_{inj}$ gives
\begin{align}
  \label{eq:interscale0} 
  C_m \frac{d}{dt} V_{m} &= I_{inj}-I_m-I_{axial},
\end{align}
where $C_m$ is the capacitance of the neuronal membrane. We have that
the gradient of the inter-compartmental leakage current $I_L$ is
\begin{equation}
  \label{eq:intercompartment_current}
  \frac{\partial I_L}{\partial x} = -I_m.
\end{equation}
The leakage current accounts for the ions that diffuse out of the
intracellular space due to the natural permeability of the neuronal
membrane.
}

\review{
Given a suitable resting potential, Ohm's law is
\begin{equation}
  I_L = G_L \frac{\partial V_m}{\partial x}.
  \label{eq:ohms_law}
\end{equation}  
Substituting $I_m$ from \eqref{eq:interscale0} using
\eqref{eq:intercompartment_current}--\eqref{eq:ohms_law}, we get what
is commonly referred to as the \emph{cable equation},
\begin{equation}
  G_{L}\frac{\partial ^{2}V_m}{\partial x^{2}} =
  C_{m}\frac{\partial V_m}{\partial t} - I_{inj} + I_{axial}.
  \label{eq:cable_equation}
\end{equation}
Given the parabolic character of this equation, the Crank-Nicholson
scheme \cite{crank_practical} is a natural choice when designing
numerical methods.
}

\review{Turning now to the compartment-discrete version,} the leakage
current in each compartment \review{$a$} is given as
\begin{align}
  \label{eq:leak}
  I^a_L = G^a_L (V^a_m-E^a_L),
\end{align}
where the membrane leakage conductivity $G^a_L$ and the leak reverse
potential $E^a_L$ are regarded as constants.

The total trans-membrane current flux now becomes
\begin{align}
  \label{eq:membraneflux}
  I^a_m = I^a_L + I^a_{ionic} = (G^a_L + \sum_T G^a_T)V^a_m - 
  (G^a_LE^a_L + \sum_T G^a_T E_T).
\end{align}

\review{In order to solve for the membrane potential in each
  compartment $a$, we start from a discrete version of
  \eqref{eq:interscale0}, using \eqref{eq:conduct} and
  \eqref{eq:membraneflux},}
\begin{align}
  \label{eq:interscale}
  C^a_m \frac{d}{dt} V^{a}_{m} &= I^a_{inj}-I^a_m-I^a_{axial}
  = I^a_{inj}-(I^a_L+\sum_T I^a_T)-I^a_{axial} \\
  \nonumber
  &= I^a_{inj}+(G^a_LE^a_L+\sum_T
  G^a_T E_T)-(G^a_L + \sum_T
  G^a_T)V^a_{m} \\
  \label{eq:membrane_ode}
  &\phantom{=} -\sum_b G^a_b V^a_{m} + \sum_b G^a_b V^b_{m}.
\end{align}

To summarize, on the intermediate scale we insert the
\review{stochastic conductivity $G^a_T$ computed from
  \eqref{eq:stochcurrent}} into \eqref{eq:ion_middle} and solve the
current-balance equation \eqref{eq:interscale} for each of the neural
compartments. We simultaneously solve for the trans-membrane current
defined in \eqref{eq:membraneflux}, to be coupled into the macroscopic
field --- next to be discussed.

\subsection{Macroscale: extracellular field potentials}
\label{sec:macroscale}

To simulate spatially non-homogeneous distributions of electrical
fields produced by single neurons and neuronal networks, we use an
electrostatic formulation of Maxwell's equations to be discretized
with finite elements. We solve for the \emph{electric field intensity}
$\fatE$ in terms of the \emph{electric scalar potential} $V$,
\begin{equation}
  \label{eq:gauge}
  \fatE = -\nabla V.
\end{equation}
The relevant dynamic form of the continuity equation with current
sources $Q_{j}$ is given by
\begin{equation}
  \label{eq:cont}
  \nabla \cdot \fatJ = -\frac{\partial \rho}{\partial t}+Q_{j},
\end{equation}
with $\fatJ$ and $\rho$ the \emph{current density} and \emph{electric
  charge density}, respectively. Further constitutive relations
include
\begin{equation}
  \label{eq:const1}
  \fatD = \varepsilon_{0} \varepsilon_{r} \fatE,
\end{equation}
and \emph{Ohm's law}
\begin{equation}
  \label{eq:const2}
  \fatJ = \sigma \fatE,
\end{equation}
in which $\fatD$ denotes the \emph{electric flux density}. Finally,
\emph{Gauss' law} states that
\begin{equation}
  \nabla \cdot \fatD = \rho.
\end{equation}
Upon taking the divergence of \eqref{eq:const2} and using the
continuity equation \eqref{eq:cont} we get
\begin{equation}
  \nabla \cdot \review{(\sigma \fatE)} = 
  -\frac{\partial \rho}{\partial t}+Q_{j}.
\end{equation}
Rewriting the electric charge density using Gauss' law together with
the constitutive relation \eqref{eq:const1} and finally applying the
gauge condition \eqref{eq:gauge} twice we arrive at the time-dependent
potential formulation
\begin{equation}
  \label{eq:form}
  -\nabla \cdot \left(\sigma \nabla V+
  \varepsilon_{0}\varepsilon_{r} \frac{\partial}{\partial t}
  \nabla V \right) = Q_{j}.
\end{equation}
The values for the electric conductivity $\sigma$ and the relative
permittivity $\varepsilon_{r}$ were obtained from
\cite{bedard_modeling_2004}. The source currents $Q_{j}$ are computed
from the compartmental model described in
\S\ref{sec:mesoscale}. Specifically, in compartment $a$, we put
\begin{align}
  \label{eq:source_currents}
  Q_j(t) &= \frac{I_m^a(t)}{A^a},
\end{align}
where $I_m^a(t)$ is obtained by solving \eqref{eq:membraneflux} and
where we recall that $A^a$ is the area of the neuronal membrane in
compartment $a$.

The boundary conditions here are homogeneous Neumann conditions
(electric isolation) everywhere except \review{in} a single point which we
take to be ground ($V = 0$). In all our simulations this point was
placed at the axis of rotation of the enclosing cylindrical
extracellular space, and directly underneath the neuronal
geometry. This procedure ensures that the formulation has a unique
solution; without this specification it is otherwise only specified up
to a constant.


\section{Numerical modeling}
\label{sec:num}

The processes at the three scales, that is, the microphysics of the
ion channel gating, the neuron current at the intermediate scale, and
the propagation of the electric field, all take place in continuous
time. Also, all processes formally affect eachother through two-way
couplings.

In \S\ref{sec:micro_meso} we discuss the numerical coupling of the ion
channel gating and the cable equation via a split-step strategy. In
\S\ref{sec:meso_macro} the propagation of the electric potential into
extracellular space is summarized and here we also explain how the
often rather complicated neuronal geometry is handled. To simplify
matters, we will disregard the usually much weaker coupling from
the external electric field back to the gating process.

\subsection{Numerical coupling of firing processes}
\label{sec:micro_meso}

To get to grip with the details of the coupling between the ion
channel gating process and the propagation of the action potential
along the neuron we need a more compact notation as follows. In
compartment $a$, denote by $\X^a(t)$ the \emph{gating state}, that is,
the number of channels being in each of the different states at time
$t$. For our sodium channel example we have $\X^a(t) =
[s_1^a,s_2^a,\ldots,s_8^a](t)^T$. The CTMC may be written compactly as
\begin{align}
  \label{eq:micro}
  d\X^a(t) &= \Stoich \fatmu^a(\X^a,V_m^a; \, dt), \\
  \intertext{for $a = 1,\ldots,\Mcompartments$, and where $\Stoich$ is
    an $\Mstates \times \Ntransitions$ matrix of integer transition
    coefficients. The dependency on the state $(\X^a,V_m^a)$ is
    implicit in the \emph{random counting measure} $\fatmu^a$
    associated with an $\Ntransitions$-dimensional Poisson process. As
    a concrete example, the transition $S_3 \longrightarrow S_4$ in
    \eqref{eq:S3toS4} satisfies}
  \label{eq:Poisson_intensity}
  \Expect[\fatmu^a_i(\X^a,V_m^a; \, dt)] &= 
  \Expect[\X^a_3 \, \alpha_m(V_m^a)] \, dt, \\
  \intertext{with}
  \Stoich_{3,i} = -\Stoich_{4,i} = -1,
\end{align}
that is, with the understanding that \eqref{eq:S3toS4} is the $i$th
transition according to some \review{given} ordering. \review{Since
  the process is a Poisson process,} \eqref{eq:Poisson_intensity}
expresses independent exponentially distributed waiting times of
intensity $\alpha_m(V_m^a)$ for each of the $\X^a_3$ possible
transitions of type $S_3 \longrightarrow S_4$.

The propagation of the action potential can similarly be written in
more compact ODE notation,
\begin{align}
  \label{eq:meso}
  dV_m^a(t) &= G(\X^a,V_m^a,\{V_m^b\}_{b \in \mathcal{B}(a)}) \, dt,
\end{align}
where $G$ is just the current-balance equation \eqref{eq:interscale}
and which depends on the state $(\X^a,V_m^a)$ via
\eqref{eq:stochcurrent}, \eqref{eq:ion_middle}, and
\eqref{eq:membraneflux}, respectively.

Eqs.~\eqref{eq:micro} and \eqref{eq:meso} form a coupled CTMC-ODE
model which falls under the scope of \emph{Piecewise Deterministic
  Markov Processes (PDMPs)} and for which numerical methods have been
investigated \cite{hybridMarkov}. A highly accurate implementation is
possible through event-detection in traditional ODE-solvers. This,
however, has a performance drawback since the ODE-solver must
continuously determine to sufficient accuracy \emph{what} micro-event
happens \emph{when}. \review{In the experiments in
  \S\ref{sec:experiments} we have chosen to rely on} a simple
split-step strategy as follows. Given a discrete time-step $\Delta t$,
and $t_{n+1} = t_n+\Delta t$,
\begin{align}
  \label{eq:micron}
  \X^a_{n+1} &= \X^a_{n}+\int_{t_n}^{t_{n+1}}
  \Stoich \fatmu^a(\X^a(s),V_{m,n}^a; \, ds), \\
  \label{eq:meson}
  V_{m,n+1}^a &= V_{m,n}^a+\int_{t_n}^{t_{n+1}}
  G(\X^a_{n+1},V_m^a(s),\{V_m^b(s)\}_{b \in \mathcal{B}(a)}) \, ds.
\end{align}
That is, \eqref{eq:micron} evolves the CTMC \eqref{eq:micro} keeping
the voltage potential fixed at its value from the previous time-step
$t_n$. Similarly, \eqref{eq:meson} evolves the ODE \eqref{eq:meso}
while keeping the state of the ion channels fixed at the time-step
$t_{n+1}$. Importantly, the usually more expensive stochastic
simulation in \eqref{eq:micron} is fully \emph{decoupled} as it only
depends on the state of the compartment $a$. The global step is
achieved separately by solving the connected ODE in \eqref{eq:meson},
and is usually quite fast.

The approximation \eqref{eq:micron}--\eqref{eq:meson} can be
understood as a split-step method and may also be analyzed as such
\cite{jsdesplit,jsdevarsplit}. The \emph{order} of the approximation
can then be expected to be $1/2$ in the root mean-square sense,
\begin{align}
  \Expect[\|\X_n-\X(t_n)\|^2+\|V_{m,n}-V_m(t_n)\|^2] \le C \Delta t.
\end{align}
Although it appears to be difficult to increase the stochastic order
of this approximation, the \emph{accuracy} is likely going to increase
by turning to symmetric Strang-type splitting methods for the
stochastic part \review{\cite{jsdesplit}}, possibly also adopting a
higher order scheme for the ODE-part. However, the efficiency of such
an approach ultimately depends on the strength of the nonlinear
feedback terms and is difficult to analyze \textit{a priori}. In the
present proof-of-concept context we aim at a convergent and consistent
coupling for which \eqref{eq:micron}--\eqref{eq:meson} are a suitable
starting point. We thus postpone the investigation of more advanced
integration methods for another occasion.

\subsection{Spatial extension of the firing process}
\label{sec:meso_macro}

The modeling in \S\ref{sec:mesoscale} did not take into account a
possible free-space potential $V_{ext}$, external to the cell. The
necessary modifications to incorporate this are as follows. For $x
\in \Omega \subset \mathbf{R}^3$, let $V_{ext}(x)$ be a given external
potential and denote by $V_{ext}^a$ the value at compartment $a$. Then
replace \eqref{eq:interscale} with
\begin{align}
  I^a_m &= I^a_L + I^a_{ionic} + I^a_{ext}  \label{eq:membraneflux2} \\
  &= (G^a_L + \sum_T G^a_T)V^a_m -
  (G^a_LE^a_L + \sum_T G^a_T E_T) + G^a_m  V^a_{ext}. \nonumber
\end{align}
for $G^a_m$ the membrane transneuronal conductivity. With this
modification, \eqref{eq:interscale} propagates the effect of the given
external field $V_{ext}(x)$ along the neuron.

In \S\ref{sec:macroscale} we discussed how the effect of a
trans-membrane current propagates into extracellular space as an
electric potential $V$
(cf.~\eqref{eq:form}--\eqref{eq:source_currents}). With the previous
discussion in mind, the following two-way coupling thus emerges: the
solution of the current-balance equation \eqref{eq:interscale} yields
a current source, which feeds into \eqref{eq:source_currents}. In
turn, this implies an external potential $V_{ext} := V$, obtained by
solving \eqref{eq:form}, finally to be inserted into
\eqref{eq:membraneflux2} above. The full model coupling thus arrived
at takes the schematic form
\begin{align}
  \mbox{\S\ref{sec:microscale}} \xrightleftharpoons{\hphantom{MM}}
  \mbox{\S\ref{sec:mesoscale}}
  \xrightleftharpoons{\hphantom{MM}} \mbox{\S\ref{sec:macroscale}}.
\end{align}
It is certainly possible to devise a numerical method using a similar
split-step strategy as in \S\ref{sec:micro_meso} for this full
coupling. \review{As mentioned before,} to simplify we shall assume
that the feedback from the external field and onto the neuron is weak,
such that $V_{ext} \approx 0$ in \eqref{eq:membraneflux2} and we thus
consider the simplified model \review{(see also
  Figure~\ref{fig:scheme_overview})}
\begin{align}
  \mbox{\S\ref{sec:microscale}} \xrightleftharpoons{\hphantom{MM}}
  \mbox{\S\ref{sec:mesoscale}}
  \xrightarrow{\hphantom{MM}} \mbox{\S\ref{sec:macroscale}}.
\end{align}
\review{This assumption is valid whenever the simulation consists of a
  small number of neighboring neurons, but becomes inaccurate if a
  large number of nearly parallel neurons is considered. See
  \cite{anastassiou_ephaptic_2011, buszaki_review_2012} for a further
  discussion.} It follows that the solution of the field potential $V$
can be done ``offline''. That is, this problem may be solved in
isolation using a pre-recorded current source $I_m^a(t)$ obtained from
the split-step method described in \S\ref{sec:micro_meso}.

The three-dimensional neuronal geometry was constructed in Comsol
Multiphysics with the help of the interface to Matlab (``LiveLink'')
by morphological additions and Boolean unification of simple geometric
objects, representing neuronal compartments. The TREES
Toolbox~\cite{cuntz2010one} was used to conveniently access the
geometrical properties of single compartments over the Matlab
interface. \review{More specifically, the geometry was constructed by
  parsing the adjacency graph $\mathcal{A}$. Starting from the initial
  node of the graph, this procedure makes sure that each compartment
  is connected to the same neighbors as in the numerical model for the
  axial current flow in \eqref{eq:conduct}.}

\begin{figure}[H]
  \centering
  \includegraphics[width=0.6\textwidth]{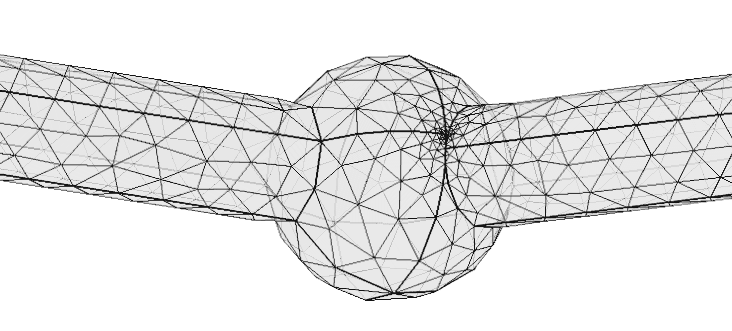}
  \caption{Example of a problematic mesh at the intersection of two
    cylindrical compartments.}
  \label{fig:mesh}
\end{figure}

In an initial attempt, we aimed to represent the 3D geometry as an
exact counterpart of the compartmental model, where each compartment
is understood as a cylinder with a certain length and diameter. If the
direction of the main axes of any two joining cylinders differed, a
sphere was added in between the cylinders, followed by a removal of
all interior boundaries. Although this created a direct volumetric
representation of the neuronal compartments, the approach is difficult
to generalize to neuronal branches with a more complicated
connectivity. The reason is that the triangulation of the final object
becomes extremely difficult to achieve as the mesh engine insists on
fully resolving the curvature. See Figure~\ref{fig:mesh} for an
example of a problematic mesh, emerging at the intersection of a
cylinder and a sphere.

\begin{figure}[H]
  \centering
  \includegraphics[width=0.45\textwidth]{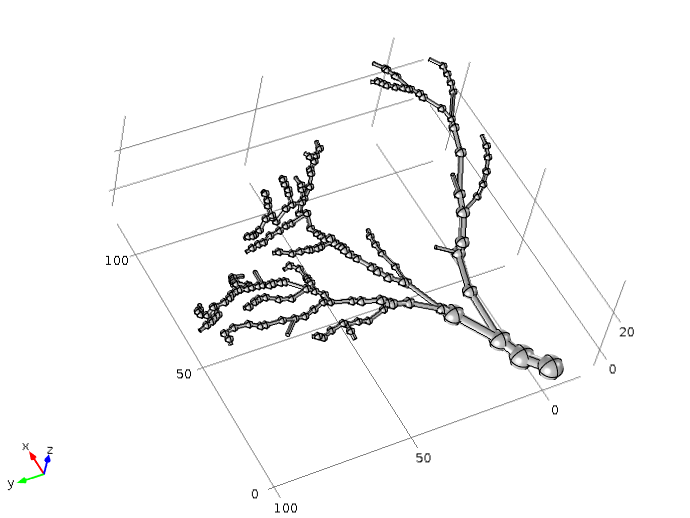}
  \includegraphics[width=0.42\textwidth]{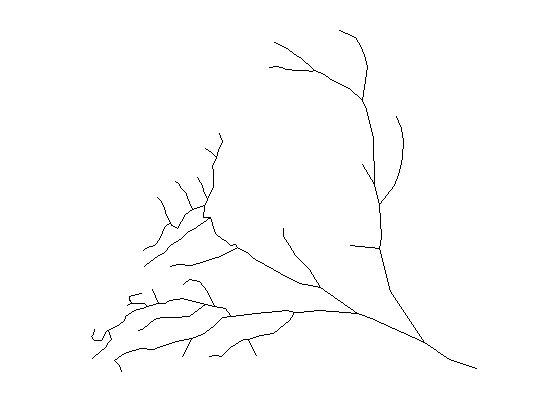}
  \caption{Example of a neuronal morphology created with cylindrical
    objects \textit{(left)}, and with curves \textit{(right)}.}
  \label{fig:morpho}
\end{figure}

A second and more successful attempt was made by which
three-dimensional curves made up of line segments for each compartment
was constructed. This simplifies the meshing process immensely, since
the extracellular mesh is not constrained by high-curvature
cylindrical boundaries. Implicit here is the assumption that the
neuron is very thin compared to the external length-scale of practical
interest. \review{This is valid as the diameter of a dendritic
  structure is $\approx 1 \mu m$ and the considered extracellular
  space is typically in the range of several $mm$
  \cite{einevoll_modelling_2013}.} In Figure~\ref{fig:morpho} we show
examples of both approaches.

\section{Experiments}
\label{sec:experiments}

We devote this section to some feasibility experiments of the method
proposed. In \S\ref{sec:micromeso_exps} we look in some detail at the
coupling of the microscopic and intermediate scales. Using the
stochastic currents so computed, the induced electrical field is
propagated outside the neuron in \S\ref{sec:full_exps}.

\subsection{Micro-meso coupling}
\label{sec:micromeso_exps}

\review{In this section we provide numerical experiments of the
  coupling between the microscopic scale, introduced in \S
  \ref{sec:microscale}, and the mesoscopic scale, introduced in \S
  \ref{sec:mesoscale}.}  We simulate the classical squid model
proposed by Hodgkin and Huxley \cite{hodgkin_quantitative_1952}, which
is also a part of a widely used benchmark for neuronal simulators
\cite{rallpackpaper}. The model includes two types of voltage-gated
ion channels, $Na^+$ and $K^+$ ions, which in our case are both
modeled as continuous-time Markov processes
(cf.~\S\ref{sec:microscale}). The kinetic gating scheme for the $Na^+$
channel is shown in Figure~\ref{fig:scheme}. The $K^+$ channel follows
a similar scheme, but has only four discrete gating states $S_1,
\ldots, S_4$, of which only the state $S_4$ is open. The transition
rates in this case are voltage-dependent as follows
\cite{hodgkin_quantitative_1952}
\begin{align}
\label{eq:rate_n_a}
\alpha_n(V_m) &= \frac{0.01(V_m+10)}{e^{(V_m+10)/10}-1},  \\
\beta_n(V_m) &= 0.125e^{(V_m/80)}.
\label{eq:rate_n_b}
\end{align}

\review{Note the similar form of \eqref{eq:rate_n_a} to
  \eqref{eq:rate1} and \eqref{eq:rate_n_b} to \eqref{eq:rate3},
  respectively, which is due to using the same empirical fitting
  procedure.}  The density of the ion channels is $\varrho_K =
30~\mu$m$^{-2}$ and $\varrho_{Na} = 330~\mu$m$^{-2}$, respectively
\cite{hille}.  The single channel conductance equals $\gamma^K = 360 /
30~pS$, and $\gamma^{Na} = 1200 / 330~pS$, while the reversal
potentials are $E_K = -77~mV$ and $E_{Na} = 50~mV$
\cite{rallpackpaper}.

The parameters for the intermediate scale are as follows. The specific
membrane capacitance $c_m = 1 \mu F (cm)^{-2}$, resting potential $E_r
= -65~mV$, cytoplasm resistivity $\varrho_c = 100~\Omega cm$, specific
leak conductance $g_L = 40000^{-1}~S(cm)^{-2}$ and leak reversal
potential $E_L = -65~mV$ \cite{rallpackpaper}. We let the model be
confined to a cylindrical geometry with a length of $1~mm$ and a
diameter of $1~\mu m$. The cylinder is compartmentalized into
$\Mcompartments$ sub-cylinders of equal length and diameter. The root
node is ignited by a current injection of $0.1~nA$.

In practice we use the Gillespie's \textit{``Direct Method''}
\cite{gillespie_exact_1977} to evolve \eqref{eq:micron} and the
Crank-Nicholson scheme \cite{crank_practical} to solve
\eqref{eq:meson}, relying on the fact that the latter is linear in
$V_m^a$.

\begin{figure}[H]
  \centering
  \includegraphics[width=1\textwidth]{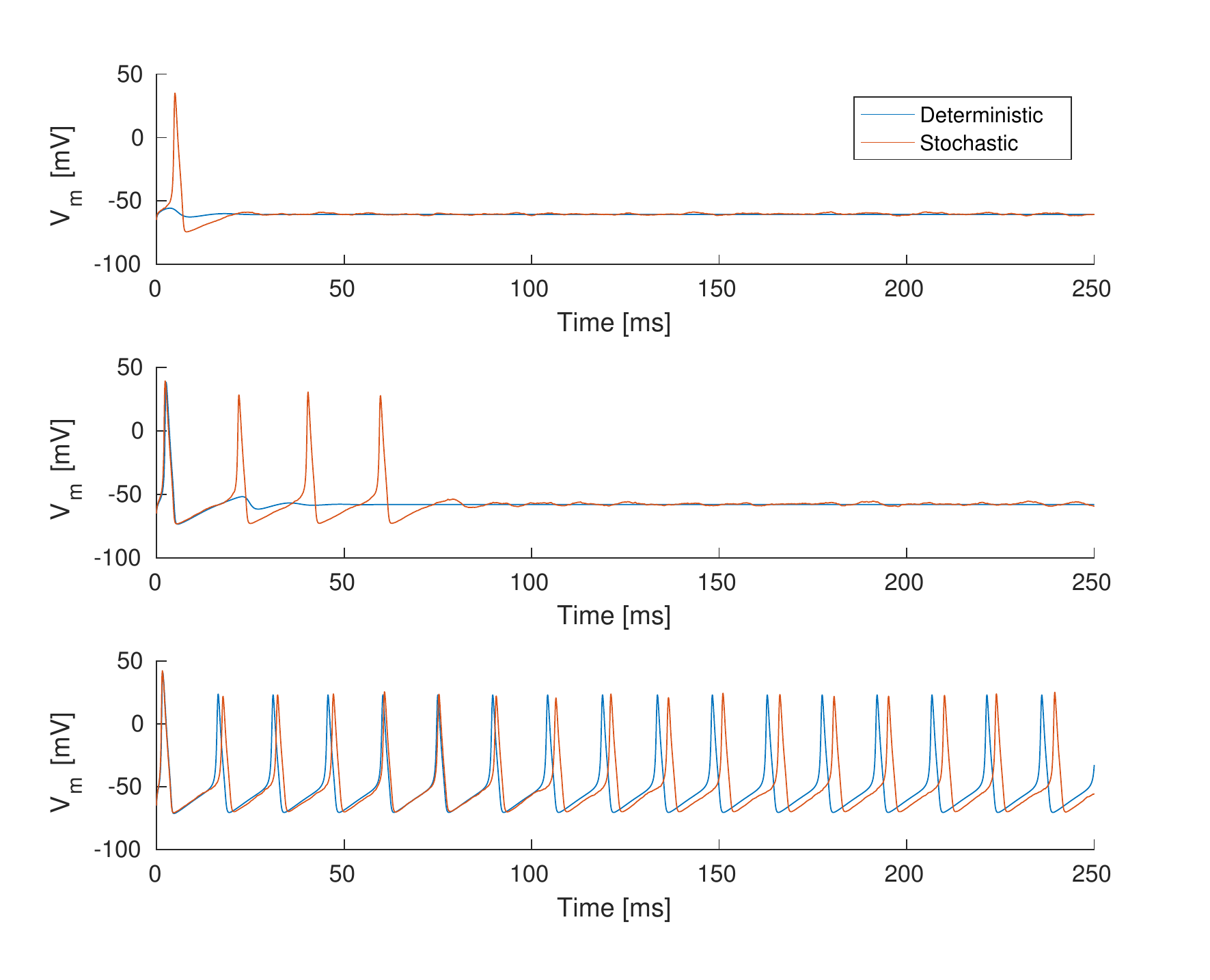}
  \caption{\review{Membrane voltage behavior around threshold
      current. The accuracy of the deterministic solver was verified
      with the reference solution of Rallpack 3. In all cases the
      discretization is $\Delta t=0.05$ and $\Delta x = 0.05$. Upper:
      injected current is $I_{inj} = 0.045$ $nA$, middle: injected
      current is $I_{inj} = 0.063$ $nA$, lower: injected current is
      $I_{inj} = 0.1$ $nA$.}}
  \label{fig:stoch_hh}
\end{figure}

\review{In Figure~\ref{fig:stoch_hh} we show the numerical solution of
  the coupled model, overlaid with the deterministic solution where
  ion channels are described by ODEs as in
  \eqref{eq:detgating1}--\eqref{eq:detgating2}.  We show three traces
  of the membrane voltage $V_m$ in one neuronal compartment over time,
  where the injected current has been varied from a lower value
  ($0.045 nA$) to a larger value ($0.1 nA$).  The dynamics of the
  stochastic model for the initial current injections clearly differ
  from the dynamics of the deterministic one, where a single spike or
  a train of spikes can be triggered in the stochastic representation,
  while no spike can be obtained in the deterministic one. At the
  higher amount of current injection, we observe a trace with similar
  characteristics for both model representations, but with an
  increasingly different phase shift.}

\begin{figure}[H]
  \centering
  \includegraphics[width=1\textwidth]{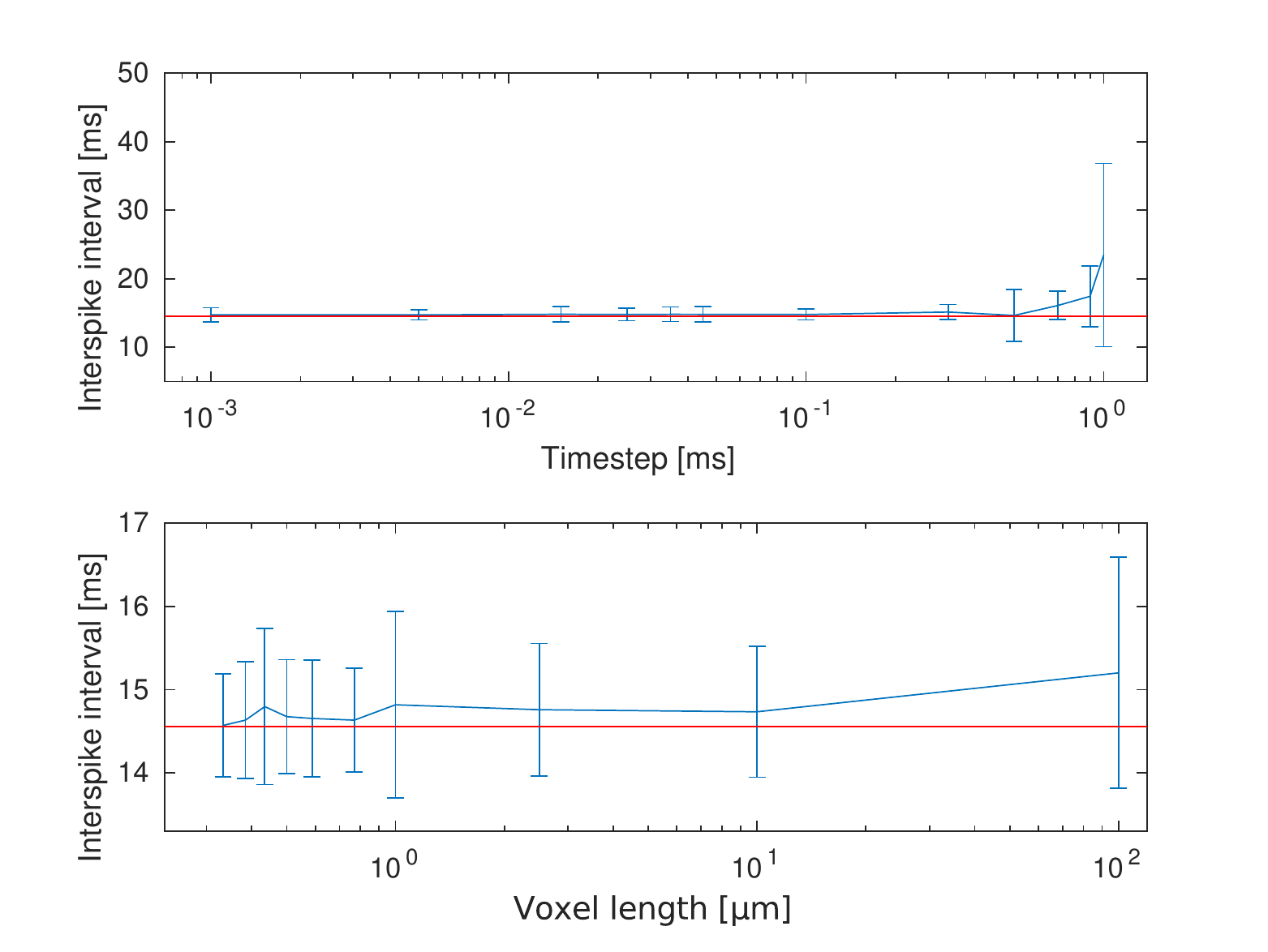}
  \caption{\review{Top: The mean interspike interval (ISI) $\pm$
      standard error as a function of the time step $\Delta t$. The
      compartment length is here $\Delta x = 0.02$. Bottom: The ISI as
      a function of the compartment length $\Delta x$ at a time step
      $\Delta t = 0.05$. The red line represents the interspike
      interval from the \textit{Rallpack 3} reference solution
      \cite{rallpackpaper}.  The number of trajectories in all runs is
      $N=40$.  }}
  \label{fig:stoch_isi}
\end{figure}

\review{In Figure~\ref{fig:stoch_isi} we show a numerical convergence
  study of the coupled model, concerning two method parameters.  We
  show how the interspike interval (ISI) changes as a function of the
  coupling time step $\Delta t$, as well as the discretization of the
  geometry $\Delta x$.  The ISI is defined as the duration between the
  peaks of two spikes. As the neuronal firing is now a stochastic
  process, the ISI will be a distribution, and hence we present the
  first and second moments. We find that, for the study of the
  coupling time step, the ISI appears to be well resolved at $\Delta t
  = 10^{-1}~ms$ for the spatial discretization presented. For the
  spatial discretization, we find that the ISI distributions do not
  significantly differ for a voxel length of under 1$\mu m$.}

\subsection{Three-scale coupling}
\label{sec:full_exps}

For this example we took the morphological description of a pyramidal
CA1 neuron from \cite{morse_abnormal_2010}, which contains about 1500
compartments. We re-sampled the geometry in order to aggregate short
compartments of sizes less than $50~\mu m$ into larger compartments,
leading to a final representation consisting of approximately 400
compartments. \review{Although the reduction might alter the
  properties of the model and more evolved protocols for compartment
  reduction could be used \cite{hendrickson_capabilities_2011,
    marasco_using_2013}, we ignore the induced discretization error
  here since the purpose of the model is to demonstrate the overall
  numerical method.}

We solved \eqref{eq:micron} and \eqref{eq:meson} over the morphology,
incorporating the active properties described in
\S\ref{sec:micromeso_exps}. Next, we scaled the transmembrane currents
$I_{m}(t)$ (cf.~\eqref{eq:source_currents}) and mapped this to the
corresponding curve segment as a current source $Q_j(t)$
\cite{ComsolACDC}. It can be noted in passing that we here assume that
transmembrane currents are the only cause of change of extracellular
potential, which is not the case in a real neuron, as for example
synaptic calcium-mediated currents are suspected to contribute to a
large fraction of the extracellular signature
\cite{buszaki_review_2012}.

Equipped with the source currents \eqref{eq:source_currents}, the
formulation \eqref{eq:form} is efficiently solved by Comsol's
\emph{``Time discrete solver''}, which is based on the observation
that the variable $W := \Delta V$ satisfies a simple ODE. Solving for
$W$ in an independent manner up to some time $t$, it is then
straightforward to solve a single static PDE to arrive at the
potential $V$ itself. For the Time discrete solver the time-step was
set to $\Delta t$ as used in the split-step method, thus ensuring a
correct transition to the macroscopic scale.

A tetrahedral mesh \cite{JinEM_FEM} was applied to discretize space
(using the ``finest'' mesh setting; resolution of curvature 0.2,
resolution of narrow regions 0.8). The simulations were verified
against coarser mesh settings in order to ensure a practically
converged solution.

\begin{figure}
  \centering

 {\includegraphics[width=35mm, angle=90]{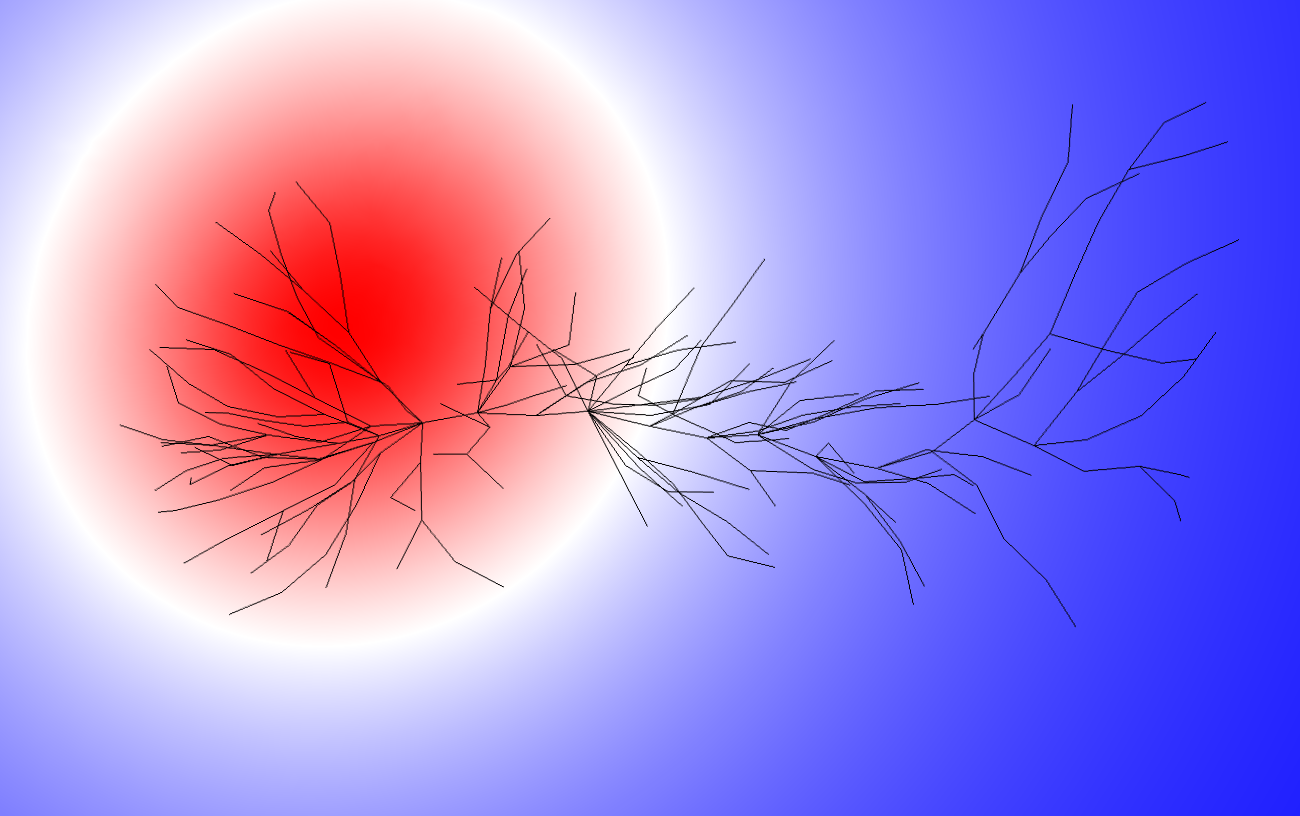}} 
 {\includegraphics[width=35mm, angle=90]{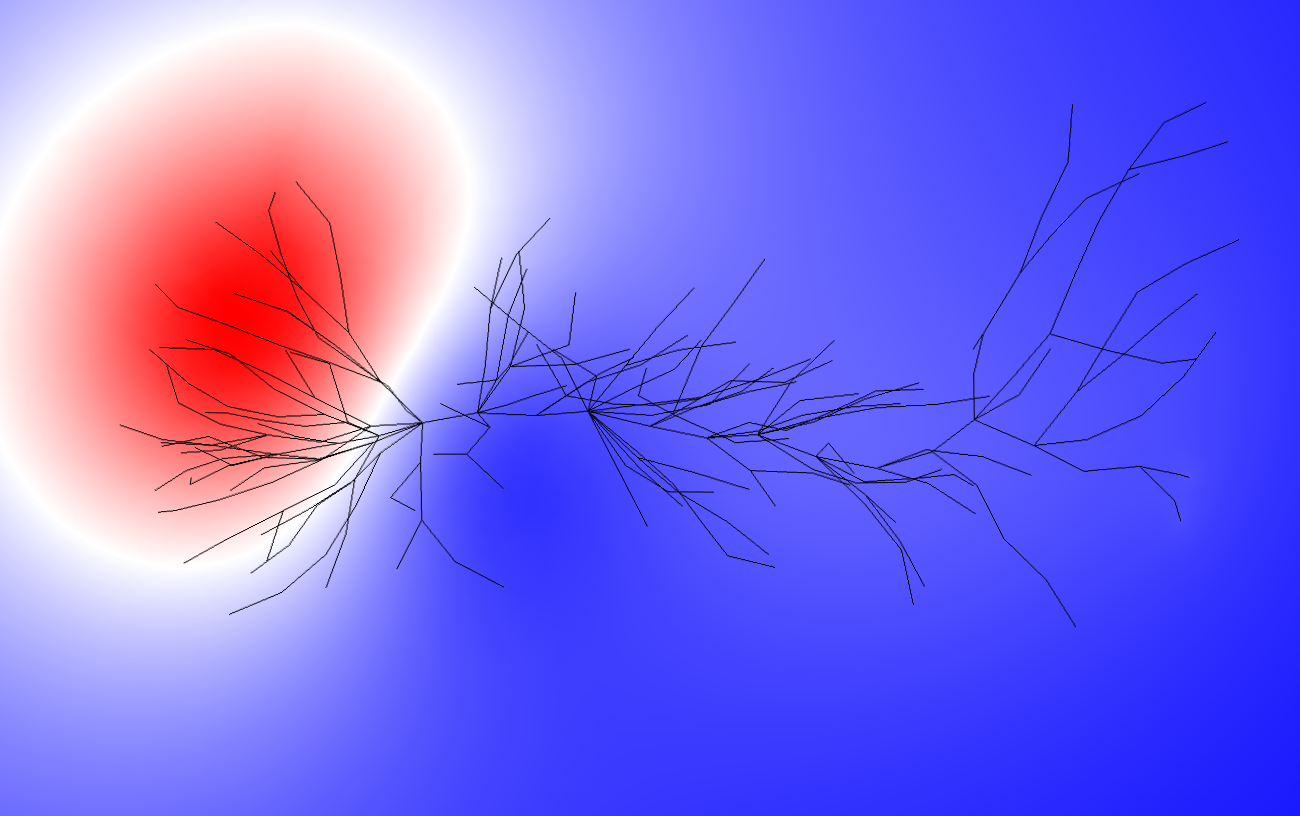}} 
 {\includegraphics[width=35mm, angle=90]{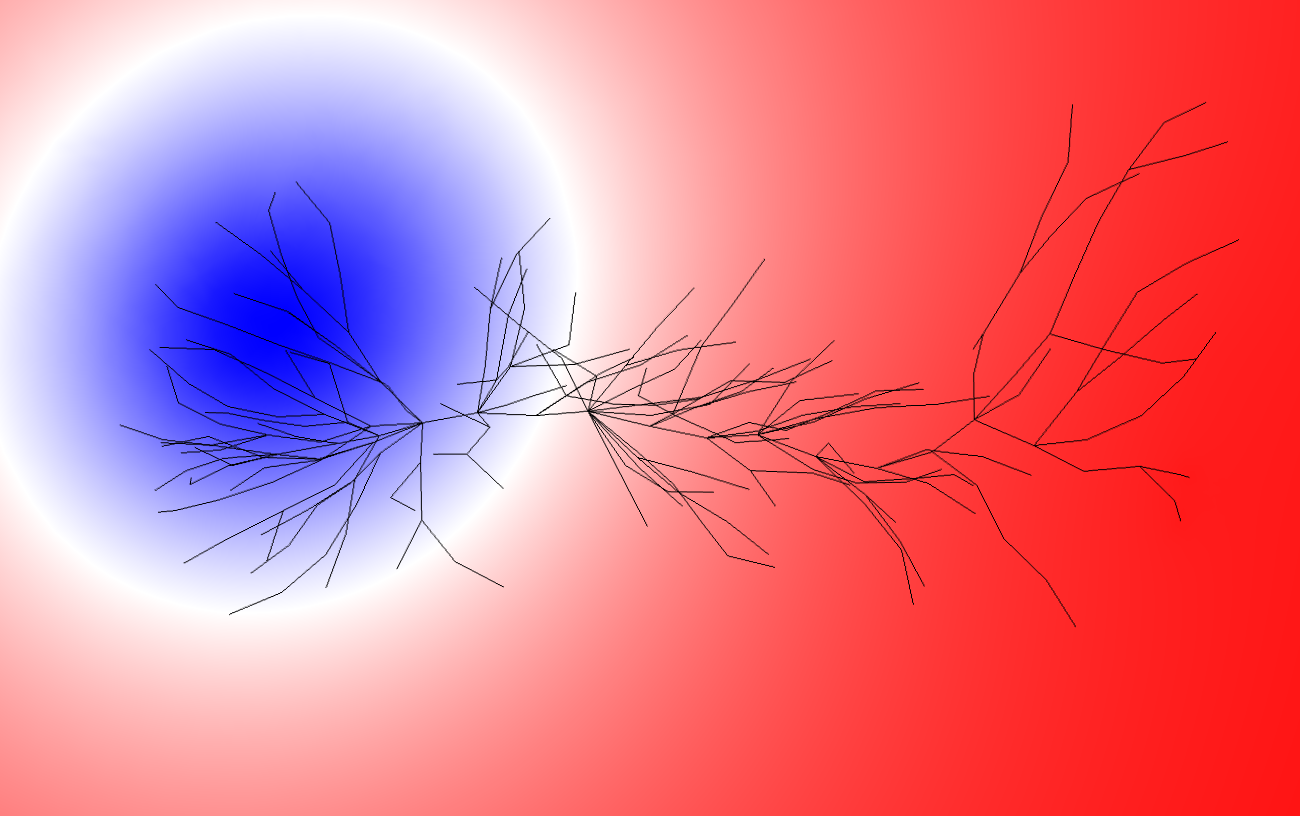}} 
 {\includegraphics[width=35mm, angle=90]{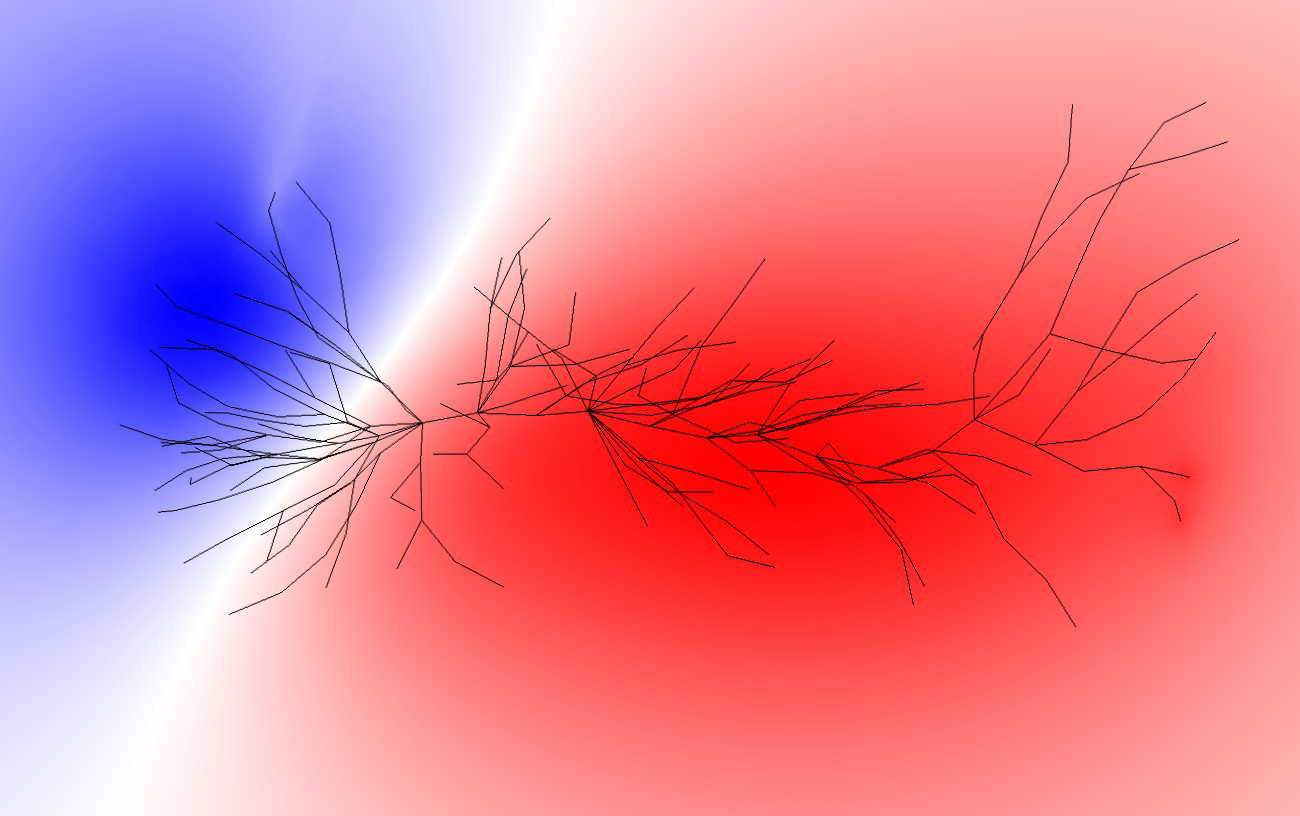}} 
 {\includegraphics[width=35mm, angle=90]{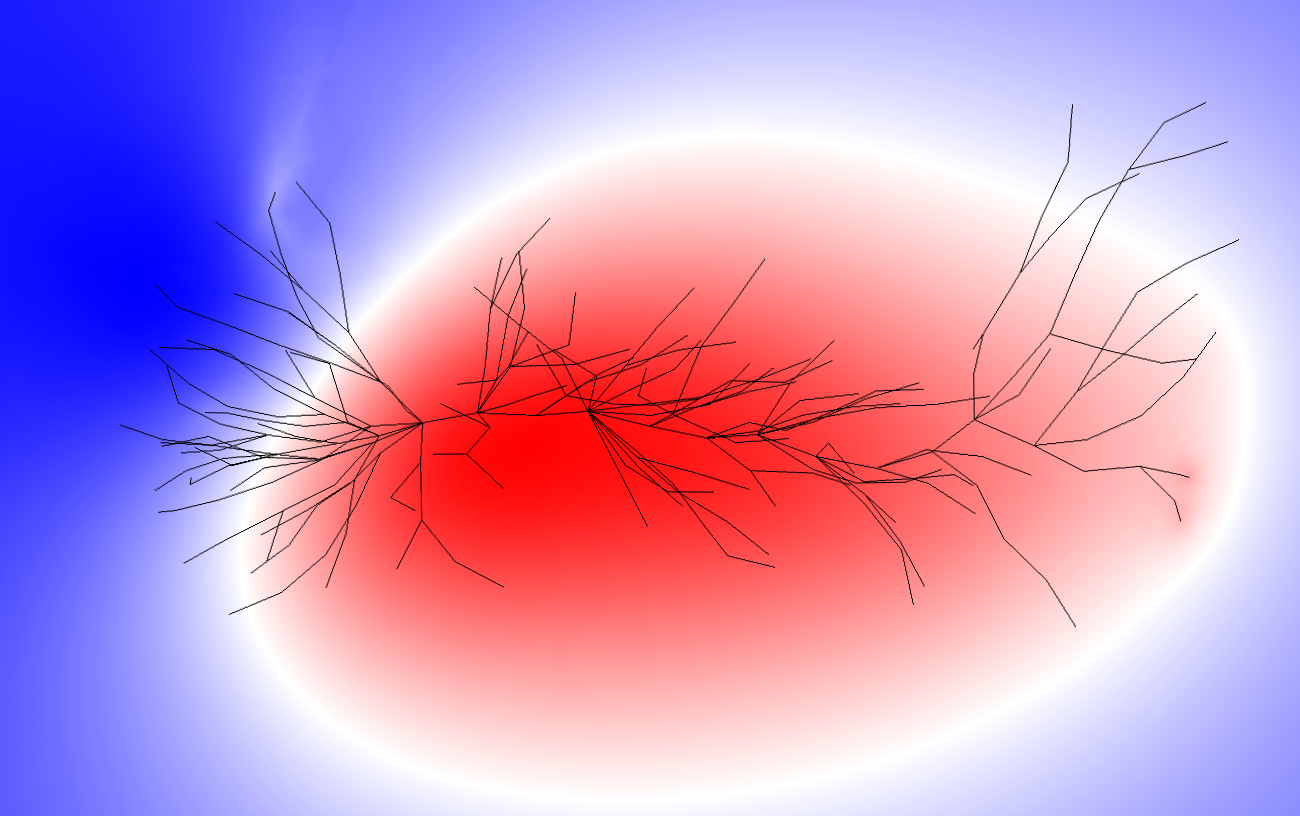}} 
    
 \caption{Solution of the full three-scale model \review{framework}:
   propagation of an extracellular action-potential into homogeneous
   extracellular space.}
 \label{fig:plane}
\end{figure}

\begin{figure}[H]
  \centering
  \includegraphics[width=0.8\textwidth]{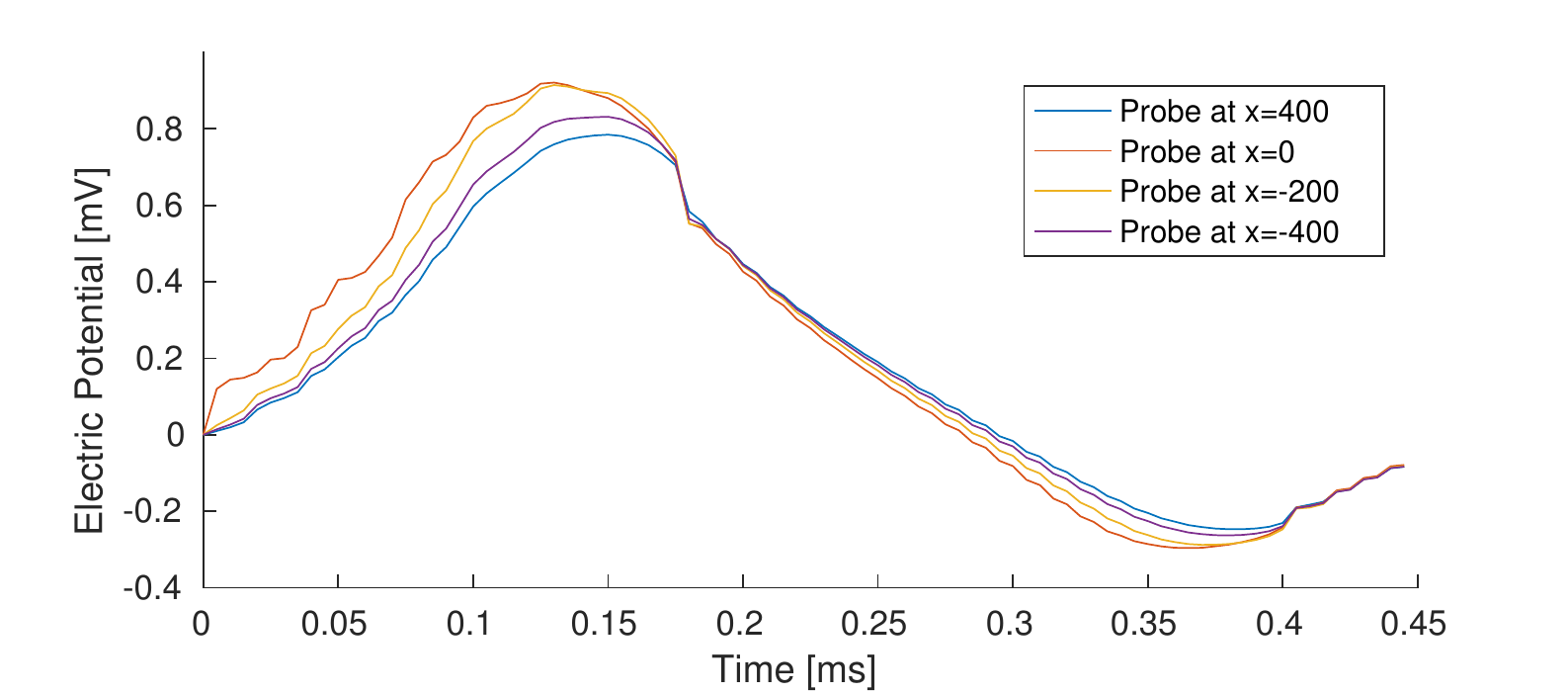}
  \caption{Electric potential (mV) in the four point probes placed
    around the neuron.}
  \label{fig:probes}
\end{figure}

The result of the simulation is visualized in Figure~\ref{fig:plane}.
We have inserted four ``point probes'' at a radial distance of
$1000~\mu m$ to the neuron, measuring the extracellular voltage at
single points. The electric potential thus monitored by the probes is
shown in Figure~\ref{fig:probes}.

\section{Conclusions}
\label{sec:conclusions}

In this paper we have proposed a model \review{framework for} neuronal
firing processes consisting of three layers: on the microscale, ion
channel gating is modeled as a continuous-time Markov chain. On the
intermediate scale, the currents produced by open channels are
integrated into the current-balance equation proposed by Hodgkin and
Huxley. Finally, on the macroscale, the outward current of neurons are
propagated into extracellular space, simulating the emission of an
extracellular potential. We have also described a numerical approach
to the coupling of the different scales and indicated through
computational results the feasibility of the overall approach.

\review{To date, several exact and approximate simulation methods
  \cite{fox_stochastic_1997, cannon_stochastic_2010,
    linaro_accurate_2011,goldwyn_stochastic_2011} of stochastically
  gated ion channel models have been proposed. However, to our
  knowledge, the coupling between the microscopic gating layer and the
  mesoscopic layer describing the action potential initiation and
  propagation, has not yet been rigorously studied. In this paper, we
  provide the formulation of the coupling as a split-step method and
  moreover numerically observe the convergence of the method with
  respect to the size of the coupling time-step as well as the spatial
  discretization. We show that the distribution of interspike
  intervals may be significantly affected by the choice of the
  coupling time-step, but does not appear to depend strongly on the
  spatial discretization. Finally, we discuss the theory and praxis of
  incorporating the macroscopic scale into the model, including the
  spatial representation of neuronal compartments and appropriate
  numerical procedures for the simulation of the local field potential
  (LFP) propagation.}

While it has often been taken for granted that ion channel gating
occurs deterministically, accumulating research evidence indicates
that the presence of stochasticity significantly influences the
neuronal behavior \cite{schneidman_ion_1998, white_channel_2000,
  faisal_noise_2008}. In turn, neurons respond with a high level of
variability to the repeated presentation of equal stimuli. This leads
to remarkable difficulties in studying the link between single cell
biophysical properties and their function in larger neuronal networks,
both in health and disease. Additionally, a recent study
\cite{cannon_stochastic_2010} has shown that stochastic ion channel
gating largely differs not only between different neuronal cell types,
but also locally between different parts of the neuron. Given the
technical difficulties of assessing the signal channel properties of
smaller neuronal compartments such as axons and dendrites, developing
reliable mathematical models is essential to tackle this problem.

\review{Several studies \cite{schneidman_ion_1998,
    white_channel_2000,faisal_noise_2008,
    cannon_stochastic_2010,stiefel_origin_2013,moezzi_ion_2016} have
  demonstrated how incorporation of channel noise into the Hodgkin and
  Huxley equations could resemble multiple realistic neuronal
  behaviors. Although it is not the scope of this work to mimic any
  particular experimental problem, but rather to provide a modeling
  framework that could be used by a diverse group of scientists in the
  future, we foresee multiple interesting applications. For example,
  it was shown previously \cite{klink_ionic_1993,
    white_noise_1998,white_channel_2000} that the stochastic nature of
  voltage-gated ion channels in the medial entorhinal cortex stellate
  cells are crucial for generating subthreshold oscillations in theta
  ($\approx 8$Hz) frequency range. These intrinsic oscillations in
  stellate cells were suggested to generate theta oscillations in vivo
  LFP \cite{white_noise_1998}. It is well established that theta
  oscillatory activity in vivo provides a temporal window in which
  spatial and declarative memories are formed
  \cite{hartley_space_2014}. Thus, our framework could be used to test
  the link between the stochastic nature of single channels in
  specific cell types and their effect at both the cellular and the
  network level.}


\section*{Acknowledgment}

PB and SE were supported by the Swedish Research Council within the
UPMARC Linnaeus center of Excellence. SM was supported by the Olle Engkvist
post-doctoral fellowship.


\section*{Availability and reproducibility}

\review{The computational results can be reproduced within the upcoming
release 1.4 of the URDME open-source simulation framework, available
for download at \url{www.urdme.org}.}


\bibliographystyle{abbrvnat}
\bibliography{mcmds}

\providecommand{\available}[1]{Available at \texttt{#1}}
\begin{thebibliography}{35}
\providecommand{\natexlab}[1]{#1}
\providecommand{\url}[1]{\texttt{#1}}
\expandafter\ifx\csname urlstyle\endcsname\relax
  \providecommand{\doi}[1]{doi: #1}\else
  \providecommand{\doi}{doi: \begingroup \urlstyle{rm}\Url}\fi

\bibitem[Anastassiou et~al.(2011)Anastassiou, Perin, Markram, and
  Koch]{anastassiou_ephaptic_2011}
C.~A. Anastassiou, R.~Perin, H.~Markram, and C.~Koch.
\newblock Ephaptic coupling of cortical neurons.
\newblock \emph{Nature Neuroscience}, 14\penalty0 (2):\penalty0 217--223, 2011.

\bibitem[B\'{e}dard et~al.(2004)B\'{e}dard, Kr\"{o}ger, and
  Destexhe]{bedard_modeling_2004}
C.~B\'{e}dard, H.~Kr\"{o}ger, and A.~Destexhe.
\newblock Modeling extracellular field potentials and the frequency-filtering
  properties of extracellular space.
\newblock \emph{Biophysical Journal}, 86\penalty0 (3):\penalty0 1829--1842,
  2004.

\bibitem[Bhalla et~al.(1992)Bhalla, Bilitch, and Bower]{rallpackpaper}
U.~S. Bhalla, D.~H. Bilitch, and J.~M. Bower.
\newblock Rallpacks: a set of benchmarks for neuronal simulators.
\newblock \emph{Trends in Neurosciences}, 15\penalty0 (11):\penalty0 453 --
  458, 1992.

\bibitem[Buzsáki et~al.(2012)Buzsáki, Anastassiou, and
  Koch]{buszaki_review_2012}
G.~Buzsáki, C.~A. Anastassiou, and C.~Koch.
\newblock The origin of extracellular fields and currents — {EEG}, {ECoG},
  {LFP} and spikes.
\newblock \emph{Nature Reviews Neuroscience}, 13\penalty0 (6):\penalty0
  407--420, 2012.

\bibitem[Cannon et~al.(2010)Cannon, O'Donnell, and
  Nolan]{cannon_stochastic_2010}
R.~C. Cannon, C.~O'Donnell, and M.~F. Nolan.
\newblock Stochastic {Ion} {Channel} {Gating} in {Dendritic} {Neurons}:
  {Morphology} {Dependence} and {Probabilistic} {Synaptic} {Activation} of
  {Dendritic} {Spikes}.
\newblock \emph{{PLoS} Computational Biology}, 6\penalty0 (8), 2010.

\bibitem[Chevallier and Engblom(2017)]{jsdevarsplit}
A.~Chevallier and S.~Engblom.
\newblock Pathwise error bounds in multiscale variable splitting methods for
  spatial stochastic kinetics, 2017.
\newblock Accepted for publication in {\it SIAM J.~Numer.~Anal.}
  \available{https://arxiv.org/abs/1607.00805}.

\bibitem[Com(2012)]{ComsolACDC}
\emph{AC/DC Module User's Guide}.
\newblock Comsol, 2012.
\newblock Version 4.3.

\bibitem[Crank and Nicolson(1947)]{crank_practical}
J.~Crank and P.~Nicolson.
\newblock A practical method for numerical evaluation of solutions of partial
  differential equations of the heat-conduction type.
\newblock \emph{Advances in Computational Mathematics}, 6\penalty0
  (1):\penalty0 207--226, 1947.

\bibitem[Cuntz et~al.(2010)Cuntz, Forstner, Borst, and Häusser]{cuntz2010one}
H.~Cuntz, F.~Forstner, A.~Borst, and M.~Häusser.
\newblock One {Rule} to {Grow} {Them} {All}: {A} {General} {Theory} of
  {Neuronal} {Branching} and {Its} {Practical} {Application}.
\newblock \emph{{PLoS} Computational Biology}, 6\penalty0 (8), 2010.

\bibitem[Dorval and White(2005)]{dorval_channel_2005}
A.~D. Dorval and J.~A. White.
\newblock Channel noise is essential for perithreshold oscillations in
  entorhinal stellate neurons.
\newblock \emph{The Journal of Neuroscience: The Official Journal of the
  Society for Neuroscience}, 25\penalty0 (43):\penalty0 10025--10028, 2005.

\bibitem[Einevoll et~al.(2010)Einevoll, W\'{o}jcik, and
  Destexhe]{einevoll_modeling_2010}
G.~T. Einevoll, D.~K. W\'{o}jcik, and A.~Destexhe.
\newblock Modeling extracellular potentials.
\newblock \emph{Journal of Computational Neuroscience}, 29\penalty0
  (3):\penalty0 367--369, 2010.

\bibitem[Einevoll et~al.(2013)Einevoll, Kayser, Logothetis, and
  Panzeri]{einevoll_modelling_2013}
G.~T. Einevoll, C.~Kayser, N.~K. Logothetis, and S.~Panzeri.
\newblock Modelling and analysis of local field potentials for studying the
  function of cortical circuits.
\newblock \emph{Nature Reviews Neuroscience}, 14\penalty0 (11):\penalty0
  770--785, 2013.

\bibitem[Engblom(2015)]{jsdesplit}
S.~Engblom.
\newblock Strong convergence for split-step methods in stochastic jump
  kinetics.
\newblock \emph{SIAM Journal of Numerical Analysis}, 53\penalty0 (6):\penalty0
  2655--2676, 2015.

\bibitem[Ethier and Kurtz(1986)]{Markovappr}
S.~N. Ethier and T.~G. Kurtz.
\newblock \emph{Markov Processes: Characterization and Convergence}.
\newblock Wiley series in Probability and Mathematical Statistics. John Wiley
  \& Sons, New York, 1986.

\bibitem[Faisal et~al.(2008)Faisal, Selen, and Wolpert]{faisal_noise_2008}
A.~A. Faisal, L.~P.~J. Selen, and D.~M. Wolpert.
\newblock Noise in the nervous system.
\newblock \emph{Nature Reviews Neuroscience}, 9\penalty0 (4):\penalty0
  292--303, 2008.

\bibitem[Fox(1997)]{fox_stochastic_1997}
R.~F. Fox.
\newblock Stochastic versions of the {Hodgkin}-{Huxley} equations.
\newblock \emph{Biophysical Journal}, 72\penalty0 (5):\penalty0 2068--2074, May
  1997.

\bibitem[Gillespie(1977)]{gillespie_exact_1977}
D.~T. Gillespie.
\newblock Exact stochastic simulation of coupled chemical reactions.
\newblock \emph{Journal of Physical Chemistry}, 81\penalty0 (25):\penalty0
  2340--2361, 1977.

\bibitem[Goldwyn et~al.(2011)Goldwyn, Imennov, Famulare, and
  Shea-Brown]{goldwyn_stochastic_2011}
J.~H. Goldwyn, N.~S. Imennov, M.~Famulare, and E.~Shea-Brown.
\newblock Stochastic differential equation models for ion channel noise in
  {Hodgkin}-{Huxley} neurons.
\newblock \emph{Physical Review. E, Statistical, Nonlinear, and Soft Matter
  Physics}, 83\penalty0 (4 Pt 1):\penalty0 041908, 2011.

\bibitem[Hartley et~al.(2014)Hartley, Lever, Burgess, and
  O'Keefe]{hartley_space_2014}
T.~Hartley, C.~Lever, N.~Burgess, and J.~O'Keefe.
\newblock Space in the brain: how the hippocampal formation supports spatial
  cognition.
\newblock \emph{Philosophical Transactions of the Royal Society of London.
  Series B, Biological Sciences}, 369\penalty0 (1635):\penalty0 20120510, 2014.

\bibitem[Hendrickson et~al.(2011)Hendrickson, Edgerton, and
  Jaeger]{hendrickson_capabilities_2011}
E.~B. Hendrickson, J.~R. Edgerton, and D.~Jaeger.
\newblock The capabilities and limitations of conductance-based compartmental
  neuron models with reduced branched or unbranched morphologies and active
  dendrites.
\newblock \emph{Journal of Computational Neuroscience}, 30\penalty0
  (2):\penalty0 301--321, 2011.

\bibitem[Hille(1992)]{hille}
B.~Hille.
\newblock \emph{{Ionic Channels of Excitable Membranes}}.
\newblock Sinauer Associates, 1992.

\bibitem[Hodgkin and Huxley(1952)]{hodgkin_quantitative_1952}
A.~L. Hodgkin and A.~F. Huxley.
\newblock A quantitative description of membrane current and its application to
  conduction and excitation in nerve.
\newblock \emph{The Journal of Physiology}, 117\penalty0 (4):\penalty0
  500--544, 1952.

\bibitem[Jin(2002)]{JinEM_FEM}
J.~Jin.
\newblock \emph{The Finite Element Method in Electromagnetics}.
\newblock John Wiley \& Sons, New York, 2nd edition, 2002.

\bibitem[Klink and Alonso(1993)]{klink_ionic_1993}
R.~Klink and A.~Alonso.
\newblock Ionic mechanisms for the subthreshold oscillations and differential
  electroresponsiveness of medial entorhinal cortex layer {II} neurons.
\newblock \emph{Journal of Neurophysiology}, 70\penalty0 (1):\penalty0
  144--157, 1993.

\bibitem[Linaro et~al.(2011)Linaro, Storace, and
  Giugliano]{linaro_accurate_2011}
D.~Linaro, M.~Storace, and M.~Giugliano.
\newblock Accurate and fast simulation of channel noise in conductance-based
  model neurons by diffusion approximation.
\newblock \emph{PLoS Computational Biology}, 7\penalty0 (3):\penalty0 e1001102,
  2011.

\bibitem[Marasco et~al.(2013)Marasco, Limongiello, and
  Migliore]{marasco_using_2013}
A.~Marasco, A.~Limongiello, and M.~Migliore.
\newblock Using {Strahler}'s analysis to reduce up to 200-fold the run time of
  realistic neuron models.
\newblock \emph{Scientific Reports}, 3:\penalty0 srep02934, 2013.

\bibitem[Moezzi et~al.(2016)Moezzi, Iannella, and McDonnell]{moezzi_ion_2016}
B.~Moezzi, N.~Iannella, and M.~D. McDonnell.
\newblock Ion channel noise can explain firing correlation in auditory nerves.
\newblock \emph{Journal of Computational Neuroscience}, 41\penalty0
  (2):\penalty0 193--206, 2016.

\bibitem[Morse et~al.(2010)Morse, Carnevale, Mutalik, Migliore, and
  Shepherd]{morse_abnormal_2010}
T.~M. Morse, N.~T. Carnevale, P.~G. Mutalik, M.~Migliore, and G.~M. Shepherd.
\newblock Abnormal {Excitability} of {Oblique} {Dendrites} {Implicated} in
  {Early} {Alzheimer}'s: {A} {Computational} {Study}.
\newblock \emph{Frontiers in Neural Circuits}, 4, 2010.

\bibitem[Purves(2012)]{purves_neuroscience_2012}
D.~Purves.
\newblock \emph{Neuroscience}.
\newblock Sinauer Associates, 2012.
\newblock 4th Edition.

\bibitem[Rees et~al.(2000)Rees, Chang, and Spencer]{rees_crystallographic_2000}
D.~C. Rees, G.~Chang, and R.~H. Spencer.
\newblock Crystallographic {Analyses} of {Ion} {Channels}: {Lessons} and
  {Challenges}.
\newblock \emph{Journal of Biological Chemistry}, 275\penalty0 (2):\penalty0
  713--716, 2000.

\bibitem[Riedler(2013)]{hybridMarkov}
M.~G. Riedler.
\newblock Almost sure convergence of numerical approximations for piecewise
  deterministic {M}arkov processes.
\newblock \emph{Journal of Computational and Applied Mathematics}, 239\penalty0
  (0):\penalty0 50--71, 2013.

\bibitem[Schneidman et~al.(1998)Schneidman, Freedman, and
  Segev]{schneidman_ion_1998}
E.~Schneidman, B.~Freedman, and I.~Segev.
\newblock Ion channel stochasticity may be critical in determining the
  reliability and precision of spike timing.
\newblock \emph{Neural Computation}, 10\penalty0 (7):\penalty0 1679--1703,
  1998.

\bibitem[Stiefel et~al.(2013)Stiefel, Englitz, and
  Sejnowski]{stiefel_origin_2013}
K.~M. Stiefel, B.~Englitz, and T.~J. Sejnowski.
\newblock Origin of intrinsic irregular firing in cortical interneurons.
\newblock \emph{Proceedings of the National Academy of Sciences of the United
  States of America}, 110\penalty0 (19):\penalty0 7886--7891, 2013.

\bibitem[White et~al.(1998)White, Klink, Alonso, and Kay]{white_noise_1998}
J.~A. White, R.~Klink, A.~Alonso, and A.~R. Kay.
\newblock Noise from voltage-gated ion channels may influence neuronal dynamics
  in the entorhinal cortex.
\newblock \emph{Journal of Neurophysiology}, 80\penalty0 (1):\penalty0
  262--269, 1998.

\bibitem[White et~al.(2000)White, Rubinstein, and Kay]{white_channel_2000}
J.~A. White, J.~T. Rubinstein, and A.~R. Kay.
\newblock Channel noise in neurons.
\newblock \emph{Trends in Neurosciences}, 23\penalty0 (3):\penalty0 131--137,
  2000.

\end{thebibliography}

\end{document}